\documentstyle[12pt,epsf,rotate]{article}

\begin{document}
\thispagestyle{empty}

\baselineskip 24pt plus 2pt minus 2pt

\hfill{DOE/ER/40561-356-INT98-00-4}

\hfill{TRI-PP-98-2}

\hfill{KRL MAP-219}
\vfill
\vspace*{16pt}
\begin{center}
{\large\bf Effective Theory for Neutron-Deuteron Scattering: Energy 
 Dependence}

\vspace*{36pt}

{\bf P.F. Bedaque$^a$}, {\bf H.-W. Hammer$^b$}, and {\bf U. van Kolck$^c$}

\vspace*{12pt}

{\sl $^a$Institute for Nuclear Theory}\\
{\sl University of Washington, Seattle, WA 98195, USA}\\
{\tt bedaque@mocha.phys.washington.edu}\\
\vspace{12pt}
{\sl $^b$TRIUMF, 4004 Wesbrook Mall }\\
{\sl Vancouver, BC, V6T 2A3, Canada}\\
{\tt hammer@alph02.triumf.ca}\\
\vspace{12pt}
{\sl $^c$ Kellogg Radiation Lab 106-38}\\
{\sl  California Institute of Technology, Pasadena, CA 91125, USA}\\
{\tt vankolck@krl.caltech.edu}

\vspace*{12pt}
\end{center}
\begin{abstract}

We report on results of the effective theory method applied to  
neutron-deuteron scattering. We extend previous results in the $J=3/2$
channel to non-zero energies and find very good agreement with experiment
without any parameter fitting.

\end{abstract}

\vfill
\newpage
\setcounter{page}{1}

Since the establishment of QCD as the theory of the strong interactions
very little progress has been made in understanding nuclear forces 
from first principles. Many phenomenological models have been 
developed with great success, 
but they all suffer from 
shortcomings, among them 
ambiguities in using nucleon-nucleon scattering information in the 
calculation of other
processes, 
difficulty in relating them to the underlying QCD, and especially, 
lack of a systematic 
expansion in a small parameter. 
The effective field theory approach has the 
promise of solving these difficulties \cite{gospel}. 
The role of the small parameter is 
played by the ratio of the typical momentum scale $Q$ in the problem to the
scale associated with the physics left out of the effective theory. 
In the case of 
nucleon interactions up to momenta of the order of 300 MeV 
one can build an effective theory containing only nucleons and pions (and delta isobars). The 
scale of the physics left out is $\sim m_\rho$ and the expansion 
parameter is 
$\sim Q/m_\rho$. This
idea was 
elaborated in a large number of works in the last few years 
\cite{wholeworld}. 
Subtle problems 
regarding the naturalness
of the shallow nuclear bound states,
renormalization,  and power counting in the presence of pion exchange 
are nowadays subject of intense discussion \cite{unendingargument,maisbira}.
However, such problems can be bypassed in those nuclear processes
where the typical momentum scale is small compared to the pion mass.
In this case one is allowed to use an effective theory without explicit pions,
contact forces being all that remains. 
That is what happens in   deuteron physics, 
since the typical momentum scale in a deuteron is given by 
the inverse of the $^3S_1$ scattering length, $1/a_t << m_\pi$. 
This situation arises because the 
nuclear potential is fine-tuned so that there is a bound state close to 
threshold with energy $\sim 1/(M a_t^2)$, much smaller than other scales in 
the problem like $\sim 1/(M r_{0t}^2)\sim m_\pi^2/M$ (we take the 
effective range 
in the $^3S_1$ channel $r_{0t}\sim 1/m_\pi$
for power counting purposes).
Attempts at model-independent
approaches in this energy range have a long history \cite{foundingfathers}.
When this approach is applied to 
nucleon-nucleon scattering up to momenta $\sim 1/a_t$ the effective range 
expansion is reproduced. The first non-trivial application is thus in the 
three-nucleon
sector. 
In this rapid communication we report results of this approach in the case 
of neutron-deuteron scattering in the $J=3/2$ channel below 
deuteron break-up.
We perform an expansion on powers of $r_{0t}/a_t$ and $p r_{0t}$, 
where $p$ is the typical momentum of the process, keeping terms up to order 
$(r_{0t}/a_t)^2,(p r_{0t})^2$ (we take $r_{0t}\sim 1/m_\pi$).
Results in extraordinary agreement for the quartet scattering length were 
previously reported in Ref. \cite{genius}. Here we extend 
this calculation
to finite energy. 

In the $J=3/2, I=1/2$ channel the spins of all three nucleons are aligned
and all two-body s-wave interactions are in the spin triplet, isospin singlet 
channel. (For this reason we will drop from now on the subscript in 
$a_t$ and $r_{0t}$).
The effective Lagrangian restricted to this channel is given by 
\cite{maisbira}

\begin{eqnarray}
{\cal L}&=&  N^\dagger(i\partial_{0}+\frac{\vec{\nabla}^{2}}{2M}+\ldots)N 
         + C_0 (N^\dagger \tau_2\vec{\sigma}\sigma_2 N)^2\\ \label{lag} 
        &+&C_2 \left[ (N^\dagger \tau_2\vec{\sigma}\sigma_2\nabla N)
               (N^\dagger \tau_2\vec{\sigma}\sigma_2\nabla N) 
            -3(N^\dagger \tau_2\vec{\sigma}\sigma_2 N)
             (N^\dagger \tau_2\vec{\sigma}\sigma_2 \nabla^2 N)+h.c.\right]
             +\ldots, \nonumber
\end{eqnarray}
where $M$ is the nucleon mass, $C_n$ are constants related to the two-body 
force terms containing $n$ derivatives, and the dots stand for higher-order
terms including relativistic corrections, 
higher-derivative terms, 
three-body forces, etc.
The constants $C_n$ are determined by nucleon-nucleon scattering data.
It turns out that, using dimensional regularization and minimal subtraction,
$C_0\sim a/M$, $C_2\sim r_0 (r_0 a)/M$, $C_4\sim r_0 (r_0 a)^2/M+\ldots$
and so on (ellipses stand for terms suppressed by powers of $r_0/a$ ).
The leading pieces in each one of these terms form a geometric series that
can be conveniently summed to all orders by the introduction of a field 
of baryon-number two \cite{transvestite}

\begin{eqnarray}
\cal L & = & N^\dagger(i\partial_{0}+\frac{\vec{\nabla}^{2}}{2M}+\ldots)N 
         + \vec{d}^\dagger\cdot(-i\partial_{0}-\frac{\vec{\nabla}^{2}}{4M}
                             +\Delta+\ldots)\vec{d} \nonumber \\ 
 &  & -\frac{g}{2} (\vec{d}^\dagger\cdot N\vec{\sigma}\sigma_2\tau_2 N 
                       +\mbox{h.c.})
 +\ldots                       \label{lagt}
\end{eqnarray}
\noindent
More generally, if the dibaryon field $\vec{d}$ is integrated out,
the Lagrangian (\ref{lag}) is recovered as long as $\Delta$ and $g$ are 
appropriate functions of $C_0$ and $C_2$. 
This resummation is by no means necessary, since for momenta of the order 
$p\sim 1/a$ the resummed terms are subleading, but it is a convenient
way of computing higher-order corrections.

\begin{figure}[htb]
\begin{center}
\epsfxsize=14cm
\centerline{\epsffile{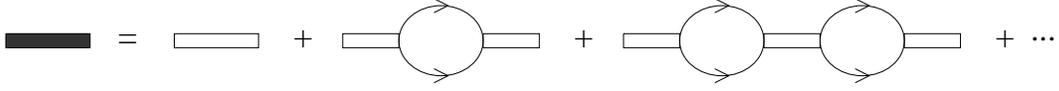}}
\end{center}
\caption{Dressed dibaryon propagator. }
\label{fig1}
\end{figure}

The numerical values of $g$ and $\Delta$ can be determined if we
consider the dressed dibaryon propagator (Fig. \ref{fig1}).
\noindent
The linearly divergent loop integral is set to zero in dimensional 
regularization
and the result is
\begin{equation}
i S(p) =  \frac {1}{p^0- \frac{\vec{p}^{\,2}}{4M} - \Delta
             + \frac{M g^{2}}{2\pi} 
               \sqrt{-M p^0+\frac{\vec{p}^{\,2}}{4}-i\epsilon} +i\epsilon} .
                                   \label{Dprop}
\end{equation}
\noindent
This propagator is, up to a constant, the scattering matrix of two nucleons
in the $^3S_1$ channel,
\begin{equation}
T(k) = {4 \pi \over M} 
                    {1\over -\frac{2 \pi \Delta}
                                  {M g^{2}} 
                    -\frac{2 \pi }
                          {M^2 g^{2}}k^2
                    -i k},      \label{NNamp}
\end{equation}
\noindent
where $k^2/M$ is the energy in the center-of-mass frame. 
This result is just the 
familiar effective range expansion, from what we can infer the proper 
values for the 
constants $g$ and $\Delta$. Using  $a= 5.42$ fm and 
$r_0=1.75$ fm \cite{nijm},
we find
\begin{eqnarray}
g^2    &= \frac{4 \pi}{M^2 r_0}  &= 1.6 \cdot 10^{-3} \ {\rm MeV^{-1}},\label{g}\\
\Delta &= \frac{2}{M a r_0}    &= 8.7 \ {\rm MeV}\label{Delta}. 
\end{eqnarray}
{}From Eqs. (\ref{Dprop}), (\ref{g}), and (\ref{Delta}) we see why it
is necessary to resum the bubble graphs in Fig. \ref{fig1} to all orders
for $p \sim 1/a$: the term in the square root coming from the unitarity
cut is of the same order as $\Delta$. On the other hand, as mentioned
before, the kinetic term of the dibaryon is smaller than the other terms
in (\ref{Dprop}) and is resummed for convenience only.
Notice that the propagator (\ref{Dprop}) has two poles, one at
$p^0= \vec{p}^{\,2}/4M-B$ (the deuteron pole), another at
 $p^0= \vec{p}^{\,2}/4M -B_{deep}$ (unphysical deep pole), 
and a cut along the positive real axis
starting at $p^0= \vec{p}^{\,2}/4M$.

\begin{figure}[htb]
\begin{center}
\epsfxsize=14cm
\centerline{\epsffile{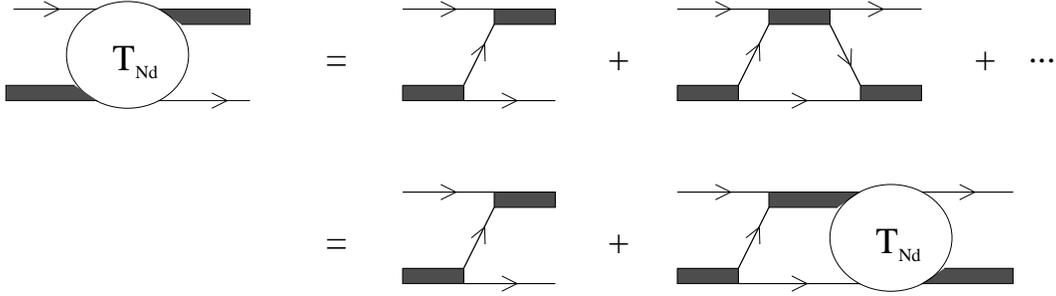}}
\end{center}
\caption{Graphs contributing up to order $(r_0/a)^2$. }
\label{fig2}
\end{figure}

 Let us now turn  to neutron-deuteron scattering. The simplest 
diagram contributing to this process is the first diagram in 
Fig. \ref{fig2}.
\noindent
{}For momenta of the order of $p \sim 1/a$ it contributes 
$\sim M g^2/p^2 \sim a^2/M r_0$. The one-loop graph
mixes different orders of the expansion, since it
involves the dibaryon propagator 
$g^2/(\Delta+p^2/M)\sim (a/M)(1+{\cal O} (r_0/a) +\ldots)$;
it gives a contribution
$\sim g^4 M^2/p \Delta \sim
(a^2/M r_0) (1 + {\cal O} (r_0/a) + \ldots)$. 
It is easy to see that the remaining graphs in 
Fig. \ref{fig2} 
give contributions of the same order, which means that an infinite number
of diagrams contribute to the leading orders. 

Other contributions are suppressed by at least three powers of 
$r_0/a$ or $p r_0$ \cite{genius}. 
For instance, the effect of the subleading (not resummed)
piece of $C_4$  is to generate
the shape parameter ($\sim k^4$) 
in the effective range expansion of the nucleon-nucleon interaction. 
Its typical size is
$\sim k^4 r_0^3$ compared to the leading piece $\sim 1/a$ and is thus also 
suppressed by $(r_0/a)^3$. 
Likewise, p-wave interactions, unaffected by the existence
of a shallow s-wave bound state, arise from a term in 
the Lagrangian with two derivatives and a coefficient of the order 
$\sim 1/M m_\pi^3$.
We conclude then that a  diagram made out of the substitution of 
one of the dibaryon 
propagators in a diagram in 
Fig. \ref{fig2} by a p-wave interaction vertex would be suppressed by
$(r_0/a)^3$ in comparison to the leading order. 
Three-body force terms have to contain at least 
two derivatives since in the $J=3/2$ channel all the spins are up and 
Fermi statistics forbids the placement of all three nucleon in a s-wave. The
natural size of the coefficient of the six nucleon, two derivative
 term that produces such a
three body force 
is $\sim 1/M m_\pi^6$. This term is generated, upon 
integration of the dibaryon field, by a term   
containing two dibaryon fields, 
two nucleon fields and two derivatives with a coefficient of the order of
 $\sim r_0^5/M a^4$. 
Thus contributions coming from 
the three-body force are suppressed in relation to the leading order 
graphs
by 
$(r_0/a)^6$.

A calculation accurate up to corrections of order $(r_0/a)^3$ is possible
by summing the diagrams of Fig. \ref{fig2}. Fortunately, the interaction 
mediated by the s-channel dibaryon generates a very simple, local and separable
potential between nucleons. It is well known that the three-body problem
with separable two-body interactions reduces to an equivalent two-body problem.
In our case the equation to be solved can be read off Fig. \ref{fig2},
and an integration over the energy inside the loop gives \cite{genius}
\begin{eqnarray}
\lefteqn{ \left[-\frac{3(\vec{p}^{\, 2}-\vec{k}^2) }
                {8 M^2 g^2}
           +\frac{1}
                 {4 \pi}({\sqrt{\frac{3}
                                          {4}(\vec{p}^{\, 2}-\vec{k}^2)+M B}
                             -\sqrt{M B}})\right]
 \frac{t(\vec{p}, \vec{k})}
      {\vec{p}^{\, 2}-\vec{k}^2-i \epsilon }}    \\ \label{aeq}
& & = \frac{-1}
           {(\vec{p}-\vec{k}/2)^2+M B}
       -   \int \frac{d^3 l}
                             {(2\pi)^3} 
       \frac{1}
            {\vec{l}^2 -\vec{l}\cdot\vec{p}+\vec{p}^{\, 2}-\frac{3}
                                                           {4}\vec{k}^2+M B}
         \frac{t(\vec{l}, \vec{k})}
              {\vec{l}^2-\vec{p}^{\, 2}-i\epsilon},  \label{fonzie}  \nonumber 
\end{eqnarray}
\noindent
where $B$ is the deuteron binding energy.
Since we are interested only in s-wave scattering, we should project
this equation into its $L=0$ component. The result is
\begin{eqnarray}
\lefteqn{  \frac{3}{2}\left[ - \eta + \frac{1}{\sqrt{\frac{3}{4}
            (x^2-y^2)+1}+1}\right]
               a(x,y)=
-\frac{1}{xy}{\rm ln}\left(\frac{(x+y/2)^2+1}{ (x-y/2)^2+1}\right)}\\
& & -\frac{2}{\pi x}\int_0^\infty \ dz\ z {\rm ln}
    \left(\frac{z^2+x^2+1 - \frac{3}{4} y^2 + xz}
               {z^2+x^2+1 - \frac{3}{4} y^2 - xz}\right) 
       \frac{a(z,y)}{z^2-y^2- i\epsilon},\nonumber 
\end{eqnarray}
\noindent
where we use the dimensionless quantities $x=p/\sqrt{MB}$, $y=k/\sqrt{MB}$, 
$z=l/\sqrt{MB}$, and 
$a(x,y)=\frac{\sqrt{ MB}}{4 \pi} 
t_{L=0}(p,k)$,
and $\eta=\sqrt{MB} r_0/2$.
For finite values of $k$ this equation is complex even below threshold 
($3 k^2/4 = B$) due to the $i\epsilon$ prescription. It is convenient for the 
numerical treatment to use the real $K$-matrix defined by
\begin{equation}
K(x,y)=\frac{a(x,y)}{1+i y a(y,y)},
\end{equation}
\noindent
which satisfies the equation
\begin{equation}
K(x,y) = - h(x,y,y) - \frac{2}{\pi}\int_0^\infty\ dz\ z^2 h(x,y,z) 
        \frac{\cal P}{z^2-y^2}K(z,y),
\end{equation}
\noindent
with
\begin{eqnarray}
h(x,y,z)  &=& \frac{1}{x z \tilde f(x,y)} {\rm ln} \left(
\frac{z^2+x^2+1-\frac{3}{4}y^2+xz}{z^2+x^2+1-\frac{3}{4}y^2-xz}
\right),\nonumber \\
\tilde f(x,y) &=&\frac{3}{2}\left[ - \eta + \frac{1}{\sqrt{\frac{3}{4}
       (x^2-y^2)+1}+1}\right].
\end{eqnarray}
The phase shifts can be obtained directly form the on-shell $K$-matrix :
\begin{equation}
k {\rm cot}\delta=\frac{\sqrt{MB}}{K(\frac{k}{\sqrt{MB}},\frac{k}{\sqrt{MB}})}.
\end{equation}
\noindent
Defining $f(x,y)$ by the equation
\begin{equation}
f(x,y)=\frac{h(x,y,y)}{h(y,y,y)}  
 -\frac{2}{\pi}\int_0^\infty dz z^2
\left( h(x,y,z)-\frac{h(x,y,y)}{h(y,y,y)}h(y,y,z)\right) 
      \frac{f(z,y)}{z^2-y^2},
\label{eqf}
\end{equation}
\noindent
the on-shell $K$-matrix can be obtained by
\begin{eqnarray}
K(y,y)& &=-h(y,y,y)\label{eqK}\\
& &\left(1+\frac{2}{\pi}\int_0^\infty dz
                     \left(z^2 h(y,y,z)f(z,y)-
                           y^2 h(y,y,y)f(y,y)\right)\frac{1}{z^2-y^2}
                 \right)^{-1}.\nonumber  
\end{eqnarray}
\noindent
Rewriting Eq. (\ref{fonzie}) this way 
greatly simplifies its numerical solution,
for now the 
integrand is regular and the principal value can be
dropped from Eqs. (\ref{eqf}) and (\ref{eqK}).

\begin{figure}[htb]
\begin{center}
\epsfxsize=10cm
\centerline{\rotate[r]{
   \rotate[r]{
             \rotate[r]{\epsffile{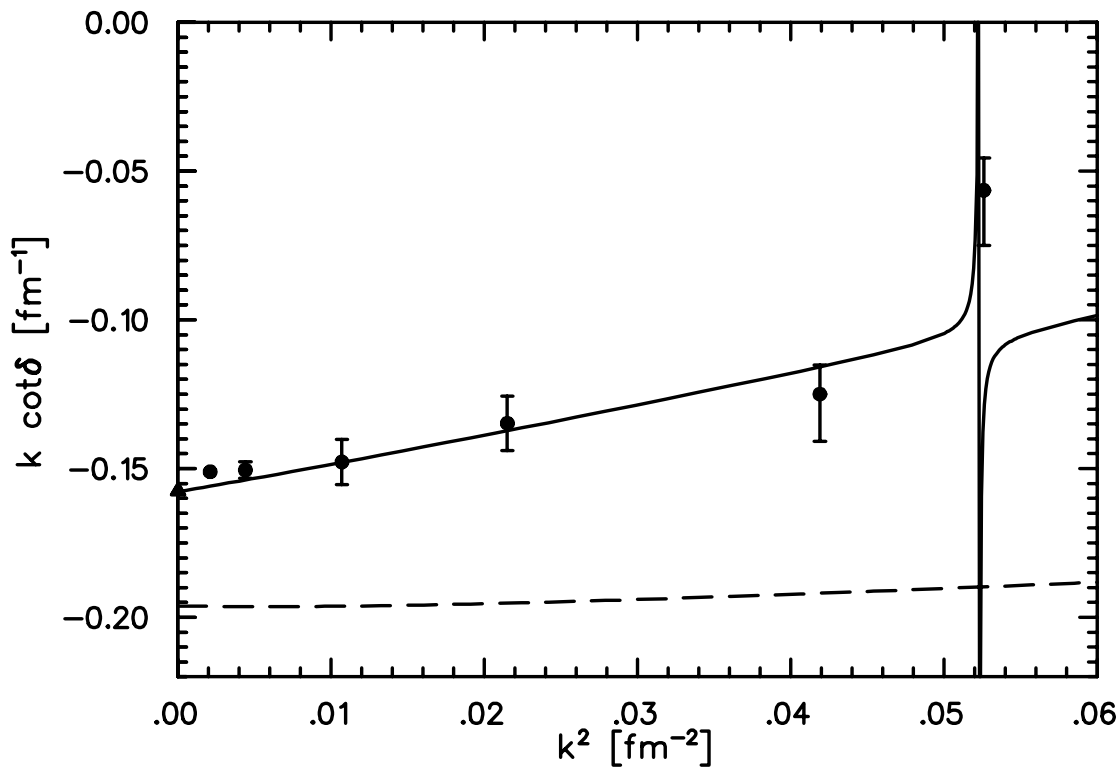}}
             }
          }}
\end{center}
\caption{$k\ {\rm cot}\delta$ in the $J=3/2$ channel to order 
$(r_0/a)^0$ (dashed line) and $(r_0/a)^2$ 
(solid line). Circles  are from the phase shift analysis in 
Ref. \protect\cite{vanOers} and the triangle is from Ref. \protect\cite{Dilg}.}
\label{fig3}
\end{figure}

We have solved Eqs. (\ref{eqf}) and (\ref{eqK}) numerically and the result for 
the phase shifts for energies up to
the break-up point is shown in 
Fig. \ref{fig3}. The data points at finite energy were taken 
from the phase shift analysis in \cite{vanOers} and the much more precise 
(nearly) zero-energy point 
from \cite{Dilg}. Also plotted is the result of the leading order calculation
obtained by setting $\eta=0$,
in which case our equations reduce to the case studied in Ref. \cite{skorny}.

We expect errors in our calculation to be of the order $(r_0/a)^3, (k r_0)^3$
compared to the leading order. These errors are smaller than the experimental
errors in the finite energy case and of the same order as the experimental 
uncertainty in the case of the more precise measurement near $k=0$, 
where we find 
$^4a_{th}=6.33\pm 0.10$ fm \cite{genius}
compared to $^4a=6.35\pm 0.02$ fm \cite{Dilg}.

Our results seem to deviate from a simple effective range type expansion 
only around the pole at $\sim 0.05$ fm$^{-2}$ . 
(A pole in $k\ {\rm cot}\delta$ corresponds to a zero in the 
scattering matrix, which does not carry any special meaning.) 
This pole does not appear in potential model calculations
({\it e.g.}, \cite{jim}), and presumably will be smoothed out
by higher-order terms that we have not yet included. 
It is interesting that the only 
``experimental'' point in this region 
seems to indicate some structure there, but 
more experimental input would be necessary to confirm the behavior  
we predict.

The calculation of higher-order corrections involves the knowledge of further
counterterms like the ones giving rise to p-wave interactions, etc. 
These
parameters can be determined either by fitting other experimental data
or by matching with another effective theory
---involving explicit pions--- valid up 
to higher energies. 
If more precise experimental data ---particularly at 
zero-energy--- appear, we would be facing
a unique situation 
where precision 
calculations in strong-interaction physics can be 
carried out \cite{yeahright...}
and tested. 

\vspace{1cm}
\noindent
{\large\bf Acknowledgments}

\noindent
We thank 
Aurel Bulgac, 
Vitaly Efimov, 
Jim Friar, 
Dirk H\"uber, David Kaplan, Willem van Oers, 
and Martin Savage for helpful discussions. 
HWH acknowledges the hospitality of the Nuclear Theory Group and
the INT in Seattle, where part of this work was carried out.  

This research was supported in part by the U.S. Department of Energy
grants DOE-ER-40561 
and DE-FG03-97ER41014, 
the Natural 
Science and Engineering Research Council of Canada, 
and the U.S. National Science Foundation grant PHY94-20470. 

\end{document}